\documentclass[preprint2]{aastex}

\shorttitle{Tidally driven dynamos in a rotating sphere}
\shortauthors{C\'ebron \& Hollerbach}

\begin{document}


\title{Tidally driven dynamos in a rotating sphere}


\author{D. C\'ebron$^{1,2}$ and R. Hollerbach$^{1,3}$}
\affil{$^1$Institut f\"ur Geophysik, Sonneggstrasse 5, ETH Z\"urich, Z\"urich, CH-8092, Switzerland}
\affil{$^2$Universit\'e Grenoble Alpes, CNRS, ISTerre, Grenoble, France}
\affil{$^3$Department of Applied Mathematics, University of Leeds, Leeds LS2 9JT, U.K.}

\email{david.cebron@ujf-grenoble.fr}
\email{r.hollerbach@leeds.ac.uk}






\begin{abstract}
   Large-scale planetary or stellar magnetic fields generated by a dynamo effect are mostly attributed to flows forced by buoyancy forces in electrically conducting fluid layers. However, these large-scale fields may also be controlled by tides, as previously suggested for the star $\tau$-boo, Mars or the Early Moon. By simulating a small local patch of a rotating fluid,  \cite{Barker2014} have recently shown that tides can drive small-scale dynamos by exciting a hydrodynamic instability, the so-called elliptical (or tidal) instability. By performing global magnetohydrodynamic simulations of a rotating spherical fluid body, we investigate if this instability can also drive the observed large-scale magnetic fields. We are thus interested by the dynamo threshold and the generated magnetic field in order to test if such a mechanism is relevant for planets and stars. Rather than solving the problem in a geometry deformed by tides, we consider a spherical fluid body and add a body force to mimic the tidal deformation in the bulk of the fluid. This allows us to use an efficient spectral code to solve the magnetohydrodynamic problem. We first compare the hydrodynamic results with theoretical asymptotic results, and numerical results obtained in a truely deformed ellipsoid, which confirms the presence of the elliptical instability. We then perform magnetohydrodynamic simulations,  and investigate the dynamo capability of the flow. Kinematic and self-consistent dynamos are finally simulated, showing that the elliptical instability is capable of generating dipole dominated large-scale magnetic field in global simulations of a fluid rotating sphere.
\end{abstract}





\section{Introduction}
 It is a commonly accepted hypothesis that buoyancy force drives planetary and stellar dynamos. Indeed, on Earth, the prevalent model is that the current magnetic field comes from thermo-chemical convective motions within the conducting liquid core. However, the validity of this convection driven dynamo model has recently been questioned in certain cases, such as Ganymede, Mercury and Mars \cite[][]{Jones2011a}, or the Early Moon \cite[][]{LeBars2011}. Alternative dynamo mechanisms, based on a different forcing, thus seem needed. However, very few natural
forcings have been identified as dynamo-capable in planets and stars:  (i) thermo-solutal convection \cite[][]{Glatzmaier_1995}, which is the standard mechanism generally applied to all planetary configurations even if it is not proved to be always relevant; and (ii) precession \cite[][]{tilgner2005precession,wu2009dynamo}, a purely mechanical forcing that
may drive dynamos \cite[][]{malkus1968precession}, despite a well-known controversy on its energetic budget (see \citealt{rochester1975can,loper1975torque} for critical discussions of this hypothesis, and \citealt{kerswell1996upper} for its rehabilitation). 

Tides have been proposed as an alternative dynamo mechanism (see e.g. \citealt{arkani2008tidal,arkani2009did} for Mars, or \citealt{LeBars2011} for the Early Moon) or as a key ingredient for the magnetic field dynamics (e.g. \citealt{Donati2008,Fares2009} for the star $\tau$ Boo), but the possibility to generate large scale magnetic fields via tides-driven flows has not been confirmed yet. Several studies \cite[e.g.][]{lacaze2006magnetic} have suggested that the flows needed for dynamo action could be provided by an elliptical (also called tidal) instability excited by tides \cite[see also][]{cebronAA}. This hydrodynamic instability, which comes from the local ellipticity $\beta$ of streamlines,  can arise in any fluid rotating at $\Omega$ provided that (i) the dimensionless fluid viscosity $E=\nu/(\Omega R^2)$ is small enough compared to $\beta$ ($\nu$ being the fluid kinematic viscosity, and $R$ the typical streamline radius), and (ii) the differential rotation between the fluid rotation and the streamlines distortion is non-zero. When this instability is present, \cite{Cebron2012b} show that the evolution of the magnetic field decay rates with the magnetic Reynolds number does not forbid the possibility of tidally driven large-scale dynamos. Recently, \cite{Barker2014} consider a small local patch of a rotating fluid, and their simulations of the elliptical instability (hereinafter abbreviated as EI) in a periodic box show that tidally driven small-scale dynamos are possible.

The present study focuses on the dynamo effect of the EI in a rotating sphere. Considering a sphere, instead of a tidally deformed geometry, allows us to benefit from the efficiency and accuracy of spectral methods. Indeed, solving the dynamo problem in a more realistic geometry, such as an ellipsoidal one, is very difficult and remains a challenge \cite[e.g.][]{Cebron2012b}. To investigate the dynamo problem we are interested in, we thus face the issue of establishing the tidal (basic) flow in global simulations of a rotating sphere, filled with an incompressible conductive fluid, and surrounded by an insulating medium. Since a conservative force has no effect in this configuration (see below for details), we can establish the tidal flow, either by imposing a boundary non-zero radial flow \cite[as in the recent work of][]{Favier2014}, or by using a non-conservative body force with non-penetrative boundary conditions. In order to avoid a fluid which suddenly becomes insulating when crossing the boundary (which would have uncertain consequences on the dynamo process), we consider here the second solution.

In section \ref{sec:intro}, we introduce the problem and the methods used in this work.  We first solve the hydrodynamic problem, in order to describe the tidally forced basic flow (section \ref{sec:basicflow}), and then the flow resulting from its destabilization by the EI when the body force is strong enough (section \ref{sec:hydro}). Then, we add in section \ref{sec:kindyn} the magnetic field into the problem, discarding its back-reaction on the flow in order to investigate the so-called kinematic dynamos. Finally, we tackle the self-consistent dynamo problem by solving the fully coupled magnetohydrodynamic equations.


\section{Description of the problem} \label{sec:intro}
We consider an incompressible Newtonian fluid of density $\rho$, kinematic viscosity $\nu$, conductivity $\sigma$, and permeability $\mu$, enclosed in a sphere of radius $R$, rotating with angular velocity $\Omega$. We choose  $R$, $\Omega^{-1}$ and $R \Omega \sqrt{\mu \rho}$ as the respective units of length, time and magnetic field. In the inertial frame of reference, the magnetic field ${\bf B}$ is governed by 
\begin{eqnarray}
  {\partial{\bf B}\over\partial t}&=& \frac{E}{Pm}\,\nabla^2{\bf B} \label{eq:induction}
  + \nabla\times({\bf u\times B}), \\
    \nabla \cdot {\bf B} &=&0, \label{eq:divB}
\end{eqnarray}
with the Ekman number $E=\nu/(\Omega R^2)$, the magnetic Prandtl number $Pm=\mu \sigma \nu$, and the fluid velocity ${\bf u}$, which is governed by
\begin{eqnarray}
{\partial{\bf u}\over\partial t} + {\bf u\cdot\nabla u} &=& -\nabla p
  + E\, \nabla^2{\bf u}+ {\bf F}_0 \nonumber \\ & & \, \, \, \, \, \, \, \, \, \, \, \,  + \{{\bf(\nabla\times B)\times B}\}, \label{eq:NS}\\
  \nabla \cdot {\bf u} &=&0, \label{eq:divu}
\end{eqnarray}
where $p=P/\rho$ is the reduced pressure, with $P$ the pressure, and the Lorentz force $\{{\bf(\nabla\times B)\times B}\}$ is removed for kinematic dynamos, and kept for self-consistent ones. The body force ${\bf F}_0$ aims at deforming the circular streamlines into elliptical ones, mimicking tidal effects. In the considered spherical geometry, the problem will be solved by taking the curl of equation (\ref{eq:NS}). To modify the flow, we thus need $\nabla \times {\bf F}_0 \neq \bf 0$, i.e. ${\bf F}_0$ cannot be a conservative force \cite[in reality, tidal forces are conservative and deform the boundary, leading to elliptical streamlines, as in][]{Cebron2012b}. We thus work with a kinematically prescribed body force, designed to drive a tidally like flow having elliptical streamlines in the bulk, i.e.
\begin{eqnarray}
{\bf F}_0=\varepsilon\, (r\sin\theta)^3\, (1-r^2)\, \cos(2\phi)\, {\bf\hat e}_s, \label{eq:force}
\end{eqnarray}
with the cylindrical radial unit vector ${\bf\hat e}_s$, and the spherical coordinates $(r,\theta,\phi)$, with the radius $r$, the colatitude $\theta$ and the azimuth $\phi$. The force amplitude $\varepsilon$ controls the amplitude of the streamlines deformation in the bulk of the fluid. Note that the regularity of $\bf u$ imposes certain constraints on the expression of  ${\bf F}_0 $  \cite[see][for details]{Lewis1990}. Note also that this force does not take into account the rotation of the tidal field due to the companion orbital motion. Indeed, the rotation of the tidal strain does not modify the physical mechanisms of the EI \cite[e.g.][]{le2010tidal}, and we thus focus on the simplest configuration, studied in detail by \cite{lacaze2004elliptical,lacaze2006magnetic}.

Equations (\ref{eq:induction})-(\ref{eq:divu}) have to be complemented with boundary conditions. Decomposing $\bf u$ as ${\bf u}=r\sin\theta\,{\bf\hat e}_\phi + {\bf u^*}$, i.e. as a solid body rotation with a unit angular velocity and a perturbation $\bf u^*$, we impose a zero angular momentum and stress-free condition for $\bf u^*$ (which is equivalent to impose a non-zero angular momentum and stress-free conditions on $\bf u$). For the magnetic field, the external region ($r>1$) is assumed to be insulating.

The problem is solved using the code H2000, described in detail in \cite{Hollerbach2000}, and \cite{Hollerbach2013}. Briefly, the code implements a pseudo-spectral method, based on a discretization with Chebychev polynomials in radius, and spherical harmonics in the angular variables. It uses the usual toroidal-poloidal decomposition, thereby automatically satisfying equations (\ref{eq:divB})  and (\ref{eq:divu}). All nonlinear products are calculated in the physical space. In this work, simulations are all performed for $E=5.10^{-3}$.

\section{Forced basic flow} \label{sec:basicflow}

   \begin{figure}[t]
   \centering
   \includegraphics[width=7cm]{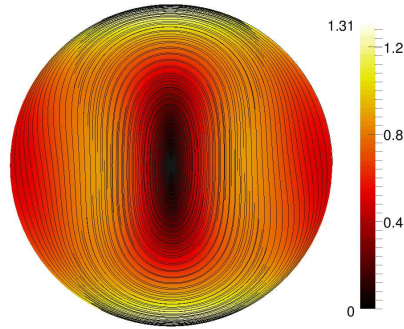}
      \caption{Velocity streamlines and magnitude $|| \mathbf{u} || $ in the equatorial plane, from a simulation of the basic flow ($\bf B=0 $,  $E=5.10^{-3}$ and $\varepsilon=10$).}
         \label{FigBasicFlow}
   \end{figure}
   
When $\varepsilon \ll 1$, the basic flow is essentially a solid body rotation, modified by the force into elliptical streamlines within the bulk of the fluid (figure \ref{FigBasicFlow}). The basic flow is thus a steady flow, which only consists of even azimuthal wavenumbers $m$, and the streamlines are smoothly deformed, from a circle at the boundary, to more and more deformed ellipses as we move towards the center. More quantitatively, noting $a$ and $b$, the long and short main axes of the elliptic streamlines, the so-called (local) streamline ellipticity $\beta=(a^2-b^2)/(a^2+b^2)$ vary from $0$, at the boundary, to a value increasing with $\varepsilon$ at the center. 

This can be analytically described for the equatorial plane in the limit $E \ll 1$, $\varepsilon \ll1$, where little algebra shows that the flow is given by the streamfunction, in polar coordinates $(s,\phi)$, as
\begin{eqnarray}
\psi=-\frac{s^2}{2}- \underbrace{\varepsilon\, \frac{(1-s^2)(5-3s^2)}{48}}_{\beta(s)}\, \frac{s^2}{2}\, \cos(2\phi), \label{eq:basicflow}
\end{eqnarray}
with the radial velocity $u_s=1/s \cdot \partial_{\phi} \psi$, and the azimuthal velocity $u_{\phi}=\partial_s \psi$. The streamlines local ellipticity $\beta(s)$ is thus $\beta=0$ at the outer boundary ($s=1$), and maximal at the center ($s=0$), where $\beta=5 \varepsilon/48 \approx0.1 \varepsilon$. This is in good agreement with the H2000 numerical basic flow, and considering for instance the one shown in figure \ref{FigBasicFlow} (where $\varepsilon$ is as large as $10$), we obtain $\beta \approx0.083 \varepsilon$ for $s=0$. 

Note that the streamline ellipticity $\beta$ is the driving parameter of the EI \cite[][]{kerswell2002elliptical,waleffe1990three}, whereas the damping parameter is $E$  \cite[stress-free conditions only leads to a volume damping, see e.g.][]{Cebron2013}. We thus expect elliptical instabilities as soon as the control parameter $\beta/E$ is large compared to $1$ somewhere in the bulk (near the center actually, where $\beta$ is maximal).

\section{Elliptical instability} \label{sec:hydro}
      \begin{figure}[!h]
   \centering
   \includegraphics[width=7cm]{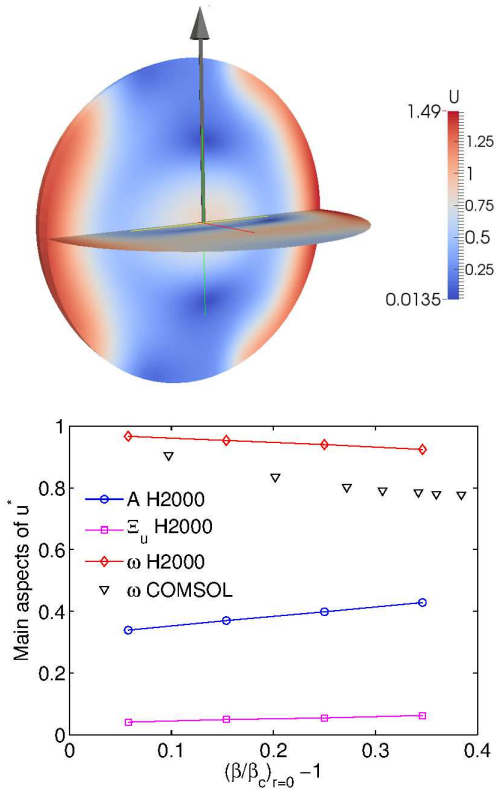}
      \caption{(top) Surface values of $\textrm{U}=||\bf{u}||$ in presence of the EI (the arrow indicates the rotation vector of the solid body rotation $r\sin\theta\,{\bf\hat e}_\phi$). (bottom) Time-average $A$, peak-to-peak value $\Xi_u$, and oscillation frequency $\omega$ of $U_{rms}$ (eq. \ref{eq:Urms}) as a function of the EI threshold distance (for $E=5.10^{-3}$, $\bf B= \bf 0 $). In the sphere, $(\beta/\beta_c)_{r=0}$ is estimated using $\varepsilon/\varepsilon_c$, which assumes that $\beta \propto \varepsilon$, as it is the case for the basic flow ($\varepsilon_c=10.4$).}
         \label{fig:comsol}
   \end{figure}

When $\varepsilon$ is above the critical value $\varepsilon_c \approx 10.4$, simulations show that the stationary basic flow is destabilized into a time-periodic flow. This instability can be investigated analytically in the limit $E \ll 1$, $\varepsilon \ll1$. Indeed, a classical (local) WKB stability analysis of the basic flow (\ref{eq:basicflow}) can be performed in this limit, by assuming the perturbations to be local plane waves characterized by their
wavevector $\mathbf{k}(t)$, with a norm $k \gg 1$, and tilted by an angle $\zeta$ to the rotation axis \cite[see e.g.][for details]{le2000three}. This analysis confirms that the basic flow (\ref{eq:basicflow}) can be destablized by an EI \cite[resonance of two plane waves for $\zeta=\pi/3$, as usual for the elliptical insability, see e.g.][]{waleffe1990three}, with a local growth rate
\begin{eqnarray}
\sigma_s= \frac{|15-72s^2+55s^4|}{256} \, \varepsilon - k^2 E,
\end{eqnarray}
which gives a maximum growth rate
\begin{eqnarray}
\sigma= \frac{15}{256} \, \varepsilon - k^2 E, \label{eq:sigWK}
\end{eqnarray}
reached for $s=0$ \cite[which is the usual inviscid growth rate $\sigma=9 \beta/16$, since $\beta=5 \varepsilon/48$ at $s=0$, see e.g.][]{waleffe1990three}. Given that the lowest $k$ is typically $k=0.5$ \cite[spinover mode, see e.g.][]{Cebron2010}, $\sigma=0$ gives a critical force amplitude of $\varepsilon_c \approx 13$ for $E=5.10^{-3}$, which is well beyond the validity condition $\varepsilon \ll 1$ of the theory, but quite close from the value $\varepsilon_c \approx 10.4$ given by the simulations.

Considering the non-linear equilibrated regime, the flow $\mathbf{u}$ is roughly an oscillating bended vortex (figure \ref{fig:comsol}, top), which oscillates approximately at the rotation rate. This time-periodic flow remains equatorially symmetric at any time (up to, at least, $\epsilon=14$), as the basic flow, but contains all azimuthal wavenumbers $m$. The flow time-evolution is actually quite complex, preventing any simple finer description. A more quantitative description of this flow is obtained by considering  the typical velocity 
\begin{eqnarray}
U_{rms}= \sqrt{\frac{2\,  E_{kin}}{V}}=  \sqrt{\frac{1}{V} \cdot \int_V \mathbf{u^*}^2\, \textrm{d} \tau} , \label{eq:Urms}
\end{eqnarray}
with $E_{kin}$ the kinetic energy of $\bf u^*$, and $V$ the volume of the fluid. In figure \ref{fig:comsol} (bottom), we show the frequency $\omega$, the time-average $A$, and the peak-to-peak value $\Xi_u$ of $U_{rms}$ as a function of the distance from the EI threshold. 
   
We can question the relevance of the flow associated with the EI driven by the force, i.e. is it possible to obtain similar flows in a truly deformed container? We have thus performed the same simulations in an ellipsoidal container, without any force ($\varepsilon=0$). To do so, we have used a numerical model, implemented in the commercial code COMSOL, which has been described and validated in detail in \cite{Cebron2013}. To be as close to a sphere as possible, the polar axis length $c$ is put equal to the long equatorial axis length $a$, and we vary the length of the small equatorial axis $b$. Similarly, an EI sets in as soon as the (uniform) ellipticity $\beta=(a^2-b^2)/(a^2+b^2)$ of the basic flow streamlines is above a certain value. Using $a=c$ as the length scale, we use the Ekman number $\nu/(\Omega a^2)=5.10^{-3}$, and vary $b$. When $\beta$ is greater than $\beta_c \approx 0.37$, an instability sets in, and the basic flow is destabilized into a periodic equatorially symmetric flow. The spatial dependency is thus the same as the flow due to EI in the sphere.

 As shown in figure \ref{fig:comsol}, the frequencies $\omega$ of the EI driven flow are comparable in the sphere and in the ellipsoid, with $\omega \approx 1$ just above threshold. This flow has been identified as the $(0,2)$ mode of the EI in \cite{Cebron2010}, i.e. a parametric resonance between the underlying strain field, and two inertial modes, with respectively $m=0$ and $m=2$.

To conclude this section on the hydrodynamic flows, the force (\ref{eq:force}) excites in the sphere an EI driven flow with the same spatial and temporal dependencies as the mode $(0,2)$ excited in an ellipsoid by the EI, which confirms the relevance of the force approach. The large amplitudes of the flow driven by the instability suggest the flow to be dynamo capable.



\section{Kinematic dynamos} \label{sec:kindyn}

To investigate kinematic dynamos, we remove the Lorentz force $\{{\bf(\nabla\times B)\times B}\}$ from the problem. This allows to study the dynamo capability of the flow studied in section \ref{sec:hydro}, without any back-reaction of $\bf B$. The goal is to identify the dynamo threshold, the critical value $Pm_c$ of $Pm$ above which we obtain exponentially growing magnetic eigenmodes.

One can first wonder if the basic flow is itself dynamo capable. Indeed, for any finite $E$, Ekman pumping leads to small axial velocities, which could a priori act as a dynamo \cite[similar dynamos have been observed by][on the Poincar\'e flow forced by precession]{tilgner2005precession}. One can also consider the flow normalized kinetic helicity 
\begin{eqnarray}
\mathcal{H}=\int_{\mathcal{V}} \, \frac{ \mathbf{u} \cdot (\nabla \times  \mathbf{u})}{ ||\mathbf{u}||\, \, ||\nabla \times  \mathbf{u}||}\, \textrm{d} \tau
\end{eqnarray}
in the south hemisphere $\mathcal{V}$, since $\mathcal{H}$ is often related to dynamo action in the literature (the flows being equatorially symmetric, $ \mathbf{u} \cdot (\nabla \times  \mathbf{u}) $ is equatorially antisymmetric, and the helicity over the whole fluid domain is thus always zero). For the basic flow, $\mathcal{H} \approx 0.04$ in the sphere ($\mathcal{H}=0$ in the ellipsoid), indicating a weak helicity. Besides, we did not find any dynamo excited on the basic flow, which does not preclude their existence at a sufficiently large $Pm$.

\begin{table}[t]
\begin{center}
\caption{Kinematic dynamos results \label{tbl-2}}
\begin{tabular}{crrrrrrrrrr}
\tableline\tableline
  & $\varepsilon=11$ & $\varepsilon=12$ & $\varepsilon=13$ & $\varepsilon=14$  \\
\tableline
$Pm_c$ &6.69 & 5.595 & 3.16  &2.47 \\
$\xi\tablenotemark{a}$ &6.69 & 6.13 & 5.69 & 5.29 \\
$\Xi_{u}$ &0.041 & 0.050 & 0.054 & 0.063 \\
\tableline
\end{tabular}
\tablenotetext{a}{With $\xi= Pm_c \cdot A({\varepsilon=11})/A$.}
\end{center}
\end{table}

For EI driven flows, kinematic dynamos are obtained, and the $Pm_c$ values (table \ref{tbl-2}) are comparable to those obtained for precession driven dynamos simulations \cite[][]{tilgner2005precession,wu2009dynamo}. Typical amplitudes of $\mathcal{H}$ in the EI equilibrated regime is $0.1$ for the sphere ($\varepsilon=11$), and $0.2$ for the deformed ellipsoid ($\beta=0.4$), showing that helicity can be generated, via the EI, in a fluid deformed by curl-free tidal forces. At $\varepsilon=14$, this amplitude is doubled in the sphere, and the decrease of $Pm_c$ with $\varepsilon$ can thus be related to this increased and rather strong helicity (however, no simple direct correlation has been found).

The decrease of $Pm_c$ with $\varepsilon$ is also partially due to the increase of the flow amplitude $A$ with $\varepsilon$. However, since $ A Pm_c/E=cst$  does not decrease as rapidly as $Pm_c$ with $\varepsilon$ (table \ref{tbl-2}), the decrease of $Pm_c$ cannot be attributed to a simple rescaling by $A$ of a single critical magnetic Reynolds number $Rm=\mu \sigma A \Omega R^2$, valid for all the unstable flows. One can thus expect $ A Pm_c/E=cst$ to be an upper bound for $Pm_c$, and writes
\begin{eqnarray}
Pm_c  \lesssim \frac{E}{A} \sim \frac{E}{\varepsilon-\varepsilon_c} \sim \frac{1}{\varepsilon/E-\alpha } , \label{eq:scaling}
\end{eqnarray}
where $\alpha=256\, k^2/15$ according to equation (\ref{eq:sigWK}). The quite large $Pm_c$ values obtained here, compared to usual real values of $Pm$ ($Pm \approx 0.01$ in the Sun's convection zone, and $Pm=10^{-5}$  in planetary cores or in liquid-metal laboratory experiments), are thus actually not a problem since these large values are due to the large Ekman number $E=5.10^{-3}$ considered here. More costly simulations, with a smaller $E$ (closer from the relevant range $E=10^{-15}-10^{-10}$ for planetary and stellar interior) will naturally lead to a smaller $\varepsilon_c$ (eq. \ref{eq:sigWK}), and thus a smaller $Pm_c$. As an example, with the typical Earth values $\beta=10^{-7}$ and $E=10^{-15}$ \cite[e.g.][]{Cebron2010}, scaling the values of table \ref{tbl-2} with equation (\ref{eq:scaling})  leads to $Pm_c \sim 10^{-5}$, with $\beta=5 \varepsilon/48$ and $k=0.5$  (see sections \ref{sec:basicflow} and \ref{sec:hydro}), showing that the relevance of this mechanism cannot be discarded on these simple arguments. For the Earth however, note that the more relevant no-slip boudary conditions do not allow to conclude about an EI excitation  \cite[see][for details, and for other planetary values of $\beta$ or $E$]{cebronAA}.

\section{Self-consistent dynamos}  \label{sec:dyndyn}
\begin{figure}[!h]
   \centering
   \includegraphics[width=7cm]{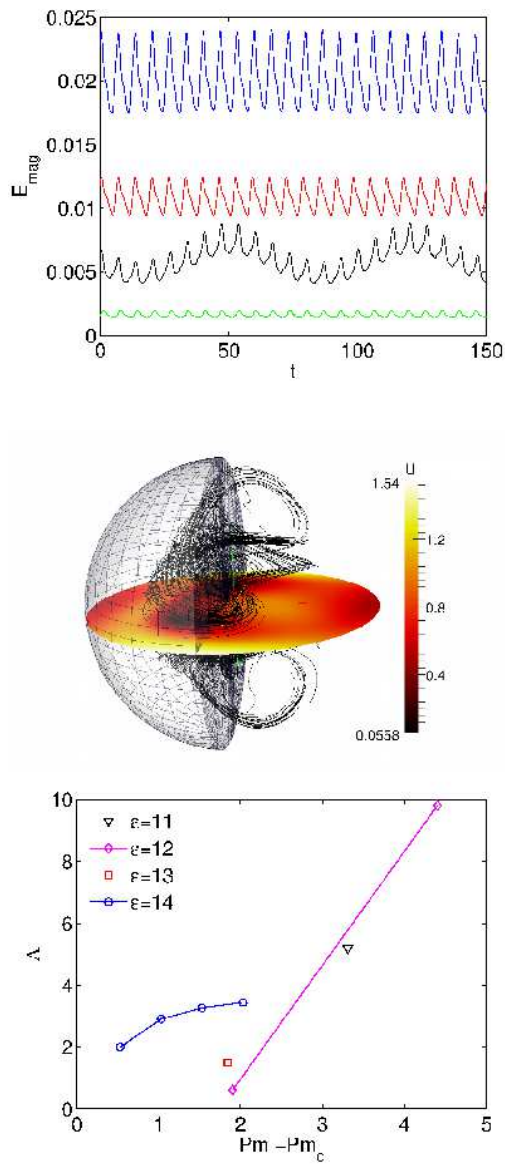}
      \caption{(top) Time evolution of the magnetic energy $E_{mag}$ for equilibrated fully coupled dynamos: from the bottom to the top, the parameters $(\varepsilon;Pm)$ are $(12;7.5)$, $(13;5)$, $(11;10)$ and $(12;10)$ .  (middle) Magnetic field lines, and velocity magnitude (equatorial slice) for the last time step ($t=150$) of the simulation $(\varepsilon;Pm)=(13;5)$ shown above. (bottom) Evolution of the Elsasser number $\Lambda$ with the dynamo threshold distance.}
         \label{FigDynT}
   \end{figure}

We next switch on the Lorentz force to obtain self-consistent dynamos. Typical time-evolutions of the magnetic energy $E_{mag}=\int_V \mathbf{B}^2\, \textrm{d} \tau$ are shown in figure \ref{FigDynT} (top), showing that the solutions can become quasi-periodic. Figure \ref{FigDynT} (middle) shows a typical snapshot of the magnetic field lines in the fluid (see below for magnetic and kinetic energies spectra).

To investigate the strength of the magnetic field generated, we define the Elsasser number $\Lambda$ as
\begin{eqnarray}
\Lambda=\frac{Pm}{E}  \frac{\left< E_{mag} \right>}{V} ,
\end{eqnarray}
which is shown in figure \ref{FigDynT} (bottom) as a function of the dynamo theshold distance $Pm-Pm_c$. We note that $\Lambda \gtrsim 1$, indicating that the Lorentz force slightly dominates the Coriolis force. For all the dynamos shown in figure \ref{FigDynT} (bottom), $B_{rms}/U_{rms}$ is between $0.05$ and $0.3$  \cite[as for the small-scale dynamos of][obtained in a periodic box]{Barker2014},and $\bf B$ oscillates at a frequency between $0.9$ and $1$, i.e. on a nearly diurnal time scale.
   \begin{figure}[t]
   \centering
   \includegraphics[width=7cm]{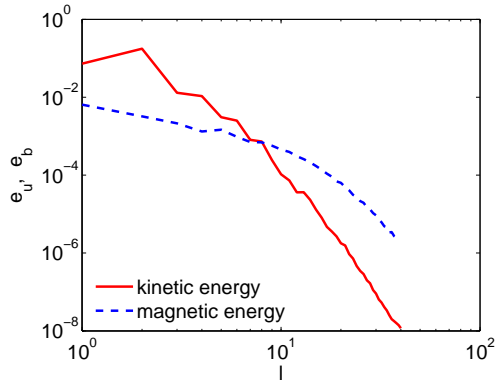}
      \caption{Spectra of time-averaged kinetic energy and magnetic energy in function of the degree $l$ of spherical harmonics ($\varepsilon=12$, $Pm=10$).}
         \label{FigSpectrum}
   \end{figure}

Figure \ref{FigSpectrum} shows time-averaged spectra of $E_{kin}$ and $E_{mag}$. The spectra shapes are actually very similar to the ones obtained for the kinematic dynamos described in section \ref{sec:kindyn}.  The magnetic field is clearly dominated by a dipolar component, without any other strongly dominant harmonics.. Note how smooth both spectra are, especially the magnetic energy. Both spectra are also clearly well-resolved. We similarly checked that the radial structure
(for which 70 Chebyshev polynomials were used) is fully resolved for each simulation of this work. We are thus fully confident that the flow driven by the EI is capable of generating dipole-dominated dynamo.

%

   \section{Conclusion}

In this work, the important CPU cost of dynamo simulations in tidally-deformed ellipsoidal domains, which requires local methods, has been avoided by using a well-designed body force in a sphere simulated using a spectral code. It has allowed to show that the elliptical instability driven in a sphere by a tidally-like force can maintain a dynamo process at magnetic Reynolds numbers comparable to critical magnetic Reynolds numbers known for other mechanical forcings. 

There is considerable further work that could be done. First, considering lower Ekman numbers would allow to use lower force amplitude, which would allow to be closer to planetary flows. Second, it would be easy to consider a rotating force, in order to excite other modes of the elliptial instability \cite[e.g.][]{le2007coriolis}. Finally, the dynamo capability of the libration driven elliptical instability, possibly excited in sychronized moons and planets \cite[e.g.][]{cebronAA,Cebron2012a}, or even of libration driven multipolar instabilities \cite[][]{Cebron2014}, could be tackled using the same approach.



\acknowledgments

This work originates from an initial idea of M. Le Bars and P. Le Gal, which has been first tackled by J. Leontini; DC is grateful to all of them for illuminating discussions about this approach. 
DC was partially supported by the ETH Z\"urich Postdoctoral fellowship Progam as well as by the Marie Curie Actions for People COFUND Program.
RH was supported in Z\"urich by European Research Council grant 247303 (MFECE),
and in Leeds by STFC grant ST/K000853/1.

\end{document}